\documentclass[aps,prb,preprint,showkeys]{revtex4}
\usepackage{amsmath,amsfonts,amssymb}
\usepackage{dsfont}
\usepackage{graphicx,subfigure,epsfig}
\usepackage{xcolor}

\newcommand{\vect}[1]{\mathbf{ #1}}
\begin{document}
\title{Spatio-temporal topology of plasmonic spin meron pairs revealed by polarimetric photo-emission microscopy}
\author{Pascal Dreher$^1$}
\author{Alexander Neuhaus$^1$}
\author{David Janoschka$^1$}
\author{Alexandra R\"odl$^1$}
\author{Tim Meiler$^2$}
\author{Bettina Frank$^2$}
\author{Timothy J. Davis$^{1,2,3}$ }
\email[Corresponding author, email: ]{timd@unimelb.edu.au}
\author{Harald Giessen$^2$ }
\email[Corresponding author, email: ]{h.giessen@pi4.uni-stuttgart.de}
\author{Frank Meyer zu Heringdorf$^1$}
\email[Corresponding author, email: ]{meyerzh@uni-due.de}
\affiliation{$^1$ Faculty of Physics and Center for Nanointegration, Duisburg-Essen (CENIDE), University of Duisburg-Essen, 47048 Duisburg, Germany}
\affiliation{$^2$ 4-th Physics Institute and Research Center SCoPE, University of Stuttgart, 70569 Stuttgart, Germany}
\affiliation{$^3$ School of Physics, University of Melbourne, Parkville Victoria 3010, Australia}

\begin{abstract}
Topology is the study of geometrical properties and spatial relations unaffected by continuous changes, and has become an important tool for understanding complex physical systems. Although recent optical experiments have inferred the existence of vector fields with the topologies of merons, the inability to extract the full three dimensional vectors misses a richer set of topologies that have not yet been fully explored. In our work, we extend the study of the topology of electromagnetic fields on surfaces to a spin quasi-particle with the topology of a meron pair, formed by interfering surface plasmon polaritons, and show that the in-plane vectors are constrained by the embedding topology of the space as dictated by the Poincar\'e-Hopf theorem. In addition we explore the time evolution of the three dimensional topology of the spin field formed by femtosecond laser pulses. These experiments are possible using our here developed method called polarimetric photo-emission electron microscopy (polarimetric PEEM) that combines an optical pump-probe technique and polarimetry with photo-emission electron microscopy. This method allows for the accurate generation of surface plasmon polariton fields and their subsequent measurement, revealing both the spatial distribution of the full three-dimensional electromagnetic fields at deep sub-wavelength resolution and their time evolution. 
\end{abstract}

\maketitle
\section*{Introduction}
The topology of spin states in matter has a profound influence on material properties, underpinning the robustness of solid state systems such as those exhibiting the quantum Hall effect. \cite{Avron2003} The topology imposes constraints on the physical parameters that stabilises these systems, as with the magnetic spins in thin films, where stable configurations often have the topology of a skyrmion.\cite{Nagaosa2013,Everschor-Sitte2018} Such vortex-like topological solitons are robust to thermal fluctuations, due to the Dzyaloshinskii–Moriya interaction that creates an energy barrier against breaking the topology. Their stability and small size holds promise for their use in information storage. \cite{Romming2013} Likewise, the half skyrmion, or meron, is an important topological feature in magnetic films where the magnetic spin vectors can be mapped onto a hemisphere. Merons are also of interest for applications in magnetic memories\cite{Knapman2021} but unlike skyrmions, merons on continuous films are not energetically stable and cannot exist in isolation. \cite{Ezawa2011, Gao2019} In these films merons tend to form in pairs but their formation in magnetic thin films is difficult to achieve and evidence for their existence has been inferred only in structured magnetic films. \cite{Shinjo2000,Phatak2012,Yu2018,Gao2019} 

In contrast to magnetic films, photonic systems provide a convenient paradigm for generating and studying vector fields with different topologies. Topological vortices associated with the vector nature of optical fields have been demonstrated in recent experiments\cite{Shen2024} but only a few have had the topology of a meron. Meron spin structures were generated in the mode structure of an optical cavity filled with liquid crystal \cite{Krol2021}, with the topology of a bimeron, and have been observed in phonon-polariton fields. \cite{Xiong2021} Merons also appear in the spin textures of interfering surface plasmon polaritons (SPPs). A trimeron spin structure \cite{Dai2020} and a meron spin lattice \cite{Ghosh2021} were inferred from measures of the in-plane electric field distributions in time-resolved photo-emission electron microscopy (PEEM). Despite these experimental demonstrations, there has been no complete measure of the spin fields of optical merons and the subsequent direct measure of their topology. Unlike magnetic films, features in the electromagnetic fields of optical systems generally have no energy barriers associated with their topology, since they are formed by interfering waves. Nevertheless, as we discuss below, there are constraints imposed by the field topology that can be revealed in their spin textures.

In this work we investigate experimentally the topology of a meron pair that we create by interfering surface plasmon polaritons (SPPs) on the surface of a gold crystal. The merons appear as a pair of spin quasi-particles arising from the rotations of the SPP electric and magnetic fields. These complex spin textures depend on the spatio-temporal properties of both the electric and magnetic field and are thus difficult to measure experimentally. We show that such complex spin textures in the SPP fields can be measured and characterized on the deep sub-wavelength scale with sub-femtosecond accuracy using a new technique that we call polarimetric PEEM, which utilises multiple distinct polarizations of the probe laser field to unambiguously extract the vector components of the meron's electric field.
From our experimental measurements we compute the Chern number that classifies the topology. We begin with a brief discussion of the topology of merons and then detail the experimental system that enables us to excite and measure the SPP electric field and to extract the complete magnetic field and spin distribution in space and time. With this information we explore the complete topology of the meron pair and show that the in-plane vectors obey the Poincar\'e-Hopf theorem that links differential geometry to topology and provides a global topological constraint on the in-plane spin vectors.

\section*{Methods}
The experiments to determine the topology of the SPP spin merons were performed using a spectroscopic photo-emission and low energy electron microscope (ELMITEC SPE-LEEM III) equipped with a TVIPS F216 detector.\cite{Janoschka2021} Optical excitation and probing of SPPs was performed using a Ti:Sapphire laser oscillator (FEMTOLASERS). The laser system generates 15 fs short pulses at a central wavelength of 800 nm and at a repetition rate of 80 MHz. The pulses are guided through the electron-optical setup of the SPE-LEEM and impinge under normal incidence.\cite{Kahl2014} For the time-resolved experiments an actively phase-stabilized Mach-Zehnder interferometer with polarization control was used (see Supplementary Note 1 for details). All polarization states were controlled using retarding waveplates and each polarization was verified using a commercial optical polarimeter. Samples were fabricated ex-situ by a single step thermolysis of (AuCl4)-tetraoctylammonium bromide on a native $\text{SiO}_2$ layer on a Si substrate.\cite{Radha2010} A focused ion beam (Raith ionLINE Plus) was used to mill grooves into the resulting Au(111) crystals. The grooves provide momentum matching and enable conversion of the laser pulses into SPPs at defined positions. After transfer to the SPE-LEEM, the samples were cleaned by standard Ar sputtering and annealing, and the work-function of the Au was lowered by deposition of a sub-monolayer of Cs. Details of the data analysis are described in Supplementary Note 2 and 3.

\section*{Results and Discussion}

The topology of a vector field is classified by the Chern number $C$, which describes the number of times the vectors map over the surface of a sphere.\cite{Gobel2021,Fosel2017} For a distribution of spin vectors $\vect s$ over the $x-y$ plane, the Chern number is obtained from a surface integral over the Chern density involving the unit spin vectors $\hat {\vect s}$,
\begin{equation}
\label{eq:1}
C=\frac{1}{4\pi} \int_A \hat {\vect{s}} \cdot \left (\partial_x \hat {\vect s} \times \partial_y \hat{ \vect s} \right) dA.
\end{equation}
With this classification, skyrmions are vector fields with a Chern number of $C=1$ whereas merons have a Chern number of $C=1/2$, which constrains the distributions of the vector fields in three dimensions. Less well-known is that the projections of vector fields onto the surface, as associated with surface plasmon polaritons, are also constrained by the two dimensional topology of the space and the distributions of the vectors on the boundary. This constraint arises from the Poincar\'e-Hopf theorem\cite{Dennis2009,Simon2018} that relates the differential geometry of the vector field to the topology of the space in which it is embedded, as represented by the Euler characteristic of the surface geometry, which is a topological invariant.

\begin{figure}[htp!]
\centering
\includegraphics[width=15.0cm]{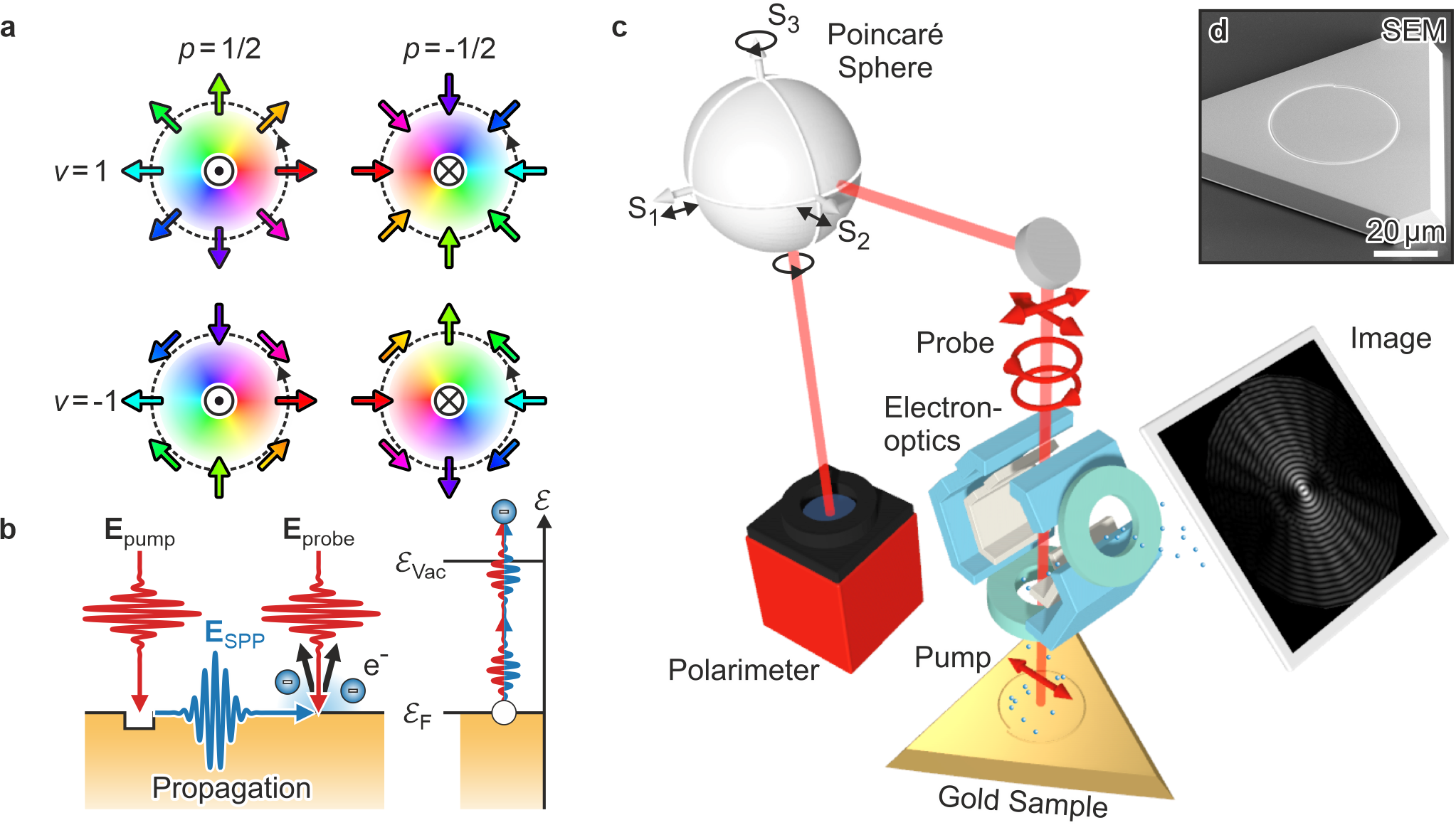}
\caption{The topology of a meron and the experimental arrangement used to measure the SPP spin texture. a) Four possible configurations of spin for a meron, represented by  $(p,v)$ with polarity $p$ and vorticity $v$. The colour hue encodes the direction of the in-plane vectors, with example arrows drawn around the boundary (black dotted line). The circles represent the direction of the out-of-plane spin (dot $s_z>0$, cross $s_z<0$). The singularity at the centre is where the in-plane vectors are zero; b) the experimental procedure involves the excitation of the SPP by the optical pump beam, the propagation of the SPP over the surface and the subsequent interference with a probe light beam, resulting in two photon absorption and the emission of a photo electron; c) a schematic of the experimental arrangement; d) a scanning electron microscope image of a single gold crystal with an Archimedean spiral etched into its surface.}
\label{fig:1}
\end{figure}

A spin texture with the topology of a meron can come in four different types related to the direction of the out-of-plane vector at its centre and the direction of rotation of the in-plane vector fields about the central point, with examples shown in Fig. \ref{fig:1}a. (Here the surface is taken as the $x-y$ plane with the normal in the $z$-direction). The point where all the in-plane spin vectors $s_x=s_y=0$ are zero is known as a C-point because the electric fields here are perfectly circularly polarised in the plane of the surface. \cite{Nye1987,Berry2001c,Fosel2017} Accordingly there is only an out-of-plane spin vector so that $s_z\neq 0$ here. The sign of this vector component determines a quantity called the polarity $p$. If the spin vector is positive $s_z>0$ then the polarity of the meron is $p=1/2$ and if the out of plane spin is negative $s_z<0$ then $p=-1/2$.\cite{Gobel2021, Shen2024} If one traverses a loop about this point in an anti-clockwise direction (black dotted line in Fig. \ref{fig:1}a), the in-plane spin vectors (depicted as arrows) rotate either in a clockwise or anti-clockwise sense. The direction of rotation of these vectors in space is known as the vorticity $v$, with $v=1$ for anti-clockwise rotation and $v=-1$ for clockwise rotation. With this characterisation the Chern number of a given isolated feature is the product \cite{Gobel2021} $C=p\cdot v$. This formula is based on the assumption that the feature has certain symmetry properties.\cite{Gobel2021} Associated with C-points are L-lines, which are lines of pure linear polarisation of the in-plane electric fields. The L-lines occur where the out-of-plane spin $s_z=0$ is zero and can be used to delineate the boundary of a local topological feature, as used in studying individual merons\cite{Dai2020} and meron lattices. \cite{Ghosh2021} 

The two-dimensional representation of the vectors in Fig. \ref{fig:1}a provides no information on the polarity of the spin, even though for this example we have arbitrarily set the polarities as shown. In principle, the sign of the polarity can be changed without affecting the vorticity. This degeneracy is a consequence of the reflection symmetry of the in-plane vectors $s_x$ and $s_y$ in the $x-y$ plane, whereas $s_z$ changes sign on reflection. \emph{As a consequence, we need to measure the complete vector field to correctly determine both the polarity and vorticity. Accordingly,  knowledge of the full three dimensional spin vectors $\vect s(x,y)$ over the surface is required to characterise the topology of any SPP spin texture that we excite}. We obtain this information using time-resolved photo-emission electron microscopy based on a new optical pump-probe method that is a significant progression of the method used to measure a skyrmion lattice created from SPP electric fields. \cite{Davis2020} 

\subsection{Polarimetric PEEM}

The SPP spin textures are generated and investigated in a pump-probe photoemission microscopy (PEEM) experiment with sub-femtosecond accuracy. The SPP spin textures are formed by illuminating a grating coupler with a femtosecond (pump) laser pulse, and the subsequent measurement is achieved by imaging the photo-electron distribution emitted by a second (probe) laser pulse in the PEEM. (Fig. \ref{fig:1}b-c). As such, the first illumination pulse excites SPPs and the second pulse probes the SPP vector field directions.\cite{Podbiel2017} The PEEM detects photo-electrons that are liberated in a second order photo-emission process (2PPE) resulting from the interfering SPP and probe electric fields\cite{Dreher2024} (Fig. \ref{fig:1}b). From data sequences created with different pump-probe time delays we reconstruct the full spatio-temporal vectorial electric field, allowing us to reconstruct the spin field. 

\begin{figure}[htp!]
\centering
\includegraphics[width=15.0cm]{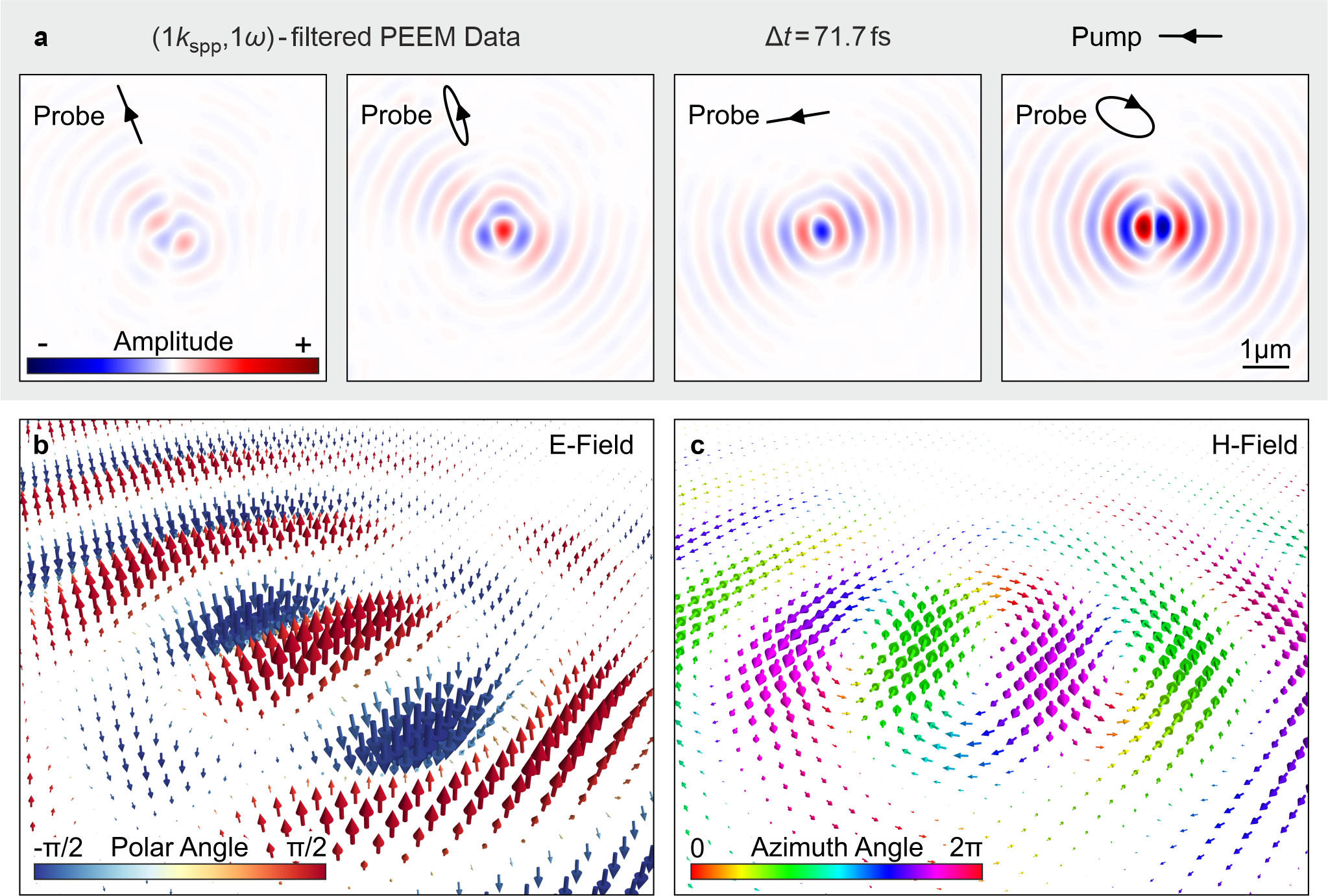}
\caption{Examples of the measured electric and magnetic fields from the polarimetric PEEM method: a) The measured SPP electric field projections $\vect E_\text{probe}^* \cdot \vect E_\text{spp}$ after Fourier filtering for four different probe polarisation states, at a pump-probe delay time of $\Delta t=71.7$ femtoseconds; b) a three dimensional rendering of the SPP electric field (time-dependency shown in Video 1) obtained by combining the data in a) and using the fact that the electric field divergence is zero; c) a three dimensional rendering of the derived magnetic field (time-dependency shown in Video 1) obtained from the electric field using Maxwell's equations. (Video 1, M4V, 16.2 MB; Video 2, M4V, 21.6 MB)}
\label{fig:2}
\end{figure}

To excite SPPs with the desired topology, the linear polarization of the first (pump) pulse is configured with respect to grooves etched into a single-crystalline gold sample (Fig. \ref{fig:1}c,d). The grooves form an Archimedean spiral that increases in radius by one SPP wavelength $\lambda_\text{spp}$ over $2\pi$ radians. See Supplementary Note 4 for details on the shape of the groove and the relative orientation of the pump polarization.  Once excited, the SPPs propagate over the surface of the metal and interfere at the center of the structure where they form vortex quasi-particles with the topology of a meron pair. The normally incident second (probe) pulse arrives after a predetermined time delay and interferes with the in-plane components of the SPP fields created by the pump pulse (Fig. \ref{fig:1}b). The combined electric field leads to a second-order absorption process and the liberation of photo-electrons with an electron yield $Y$ related to the square of the time-integrated intensity $I$
\begin{equation}
\label{eq:yield1}
\begin{split}
Y\propto \int I^2 \text{d}t &\propto \int|E_\text{probe}|^4+|E_\text{spp}|^4+2|E_\text{probe}|^2|E_\text{spp}|^2 +4 \text{Re}\left ( \vect E_\text{probe}^* \cdot \vect E_\text{spp} \right )^2 \\
&+4 \left (|E_\text{probe}|^2 +|E_\text{spp}|^2 \right ) \text{Re}\left ( \vect E_\text{probe}^* \cdot \vect E_\text{spp} \right ) \text{d}t,
\end{split}
\end{equation}
with the probe and SPP electric fields given by $\vect E_\text{probe}$ and $\vect E_\text{spp}$ respectively. The integral accounts for the time integration by the electron detector of the PEEM. The last term in Eq. \eqref{eq:yield1} yields information on the projection of the in-plane SPP electric field with the probe field via $\vect E_\text{probe}^* \cdot \vect E_\text{spp}$ and only this term contains the  fundamental  SPP wavevector $\vect k_\text{spp}$ and the fundamental frequency $\omega$  (see Supplementary Note 3 for a more detailed discussion of the scaling of the terms). We isolate this term from the extraneous ones in Eq. \eqref{eq:yield1} by performing a spatio-temporal Fourier decomposition of the photo-electron yield \cite{Davis2020,Podbiel2017}. Repeating the experiment with different probe polarizations and extracting these signal components enables the full reconstruction of the SPP vector field. Examples of PEEM measurements after Fourier filtering for the four different probe polarisations are shown in Fig.\ref{fig:2}a.

There are four key issues with reconstructing the electric field vectors.  Firstly, the polarisation of the probe pulse needs to be accurately measured, which is done before each experiment to  determine the Stokes vectors that characterise the precise polarisation. Moreover, we use four different polarisation states (two linear and two elliptical) and apply a minimisation procedure to combine the results to improve the measurement statistics. Secondly, the images from each measurement must be aligned to correct for thermal drift of the sample in the PEEM vacuum chamber, otherwise the field vectors cannot be accurately reconstructed. Thirdly, identifying the time zero, when the pump and probe pulse overlap exactly, is necessary for all the probe pulses in order to correctly phase the different images to obtain accurate measures of the SPP field vectors and their time variation. This is achieved using a spectral interference method, described in the Supplementary Note 2. Finally, because of the difficulty in centering the focused pump beam on the sample in the PEEM vacuum chamber, the Archimedean spiral is not exactly uniformly illuminated, which results in additional terms to the SPP field measurement. However, these artefacts have an asymmetric spatial dependence that can be removed by symmetrizing the fields in the Fourier space. The field reconstruction is then performed by a fixed-point iteration, as discussed in detail in Supplementary Note 3.  The result of the reconstruction is a very clean measure of the SPP field $\vect{E}_\text{spp}(\vect r,t)$ as a function of position $\vect r$ over the sample surface and as a function of time $t$. 
  
With this \emph{polarimetric} PEEM technique, we obtain the time evolution of the SPP electric field $\vect E_\text{spp}(\vect r,t)e^{-i\omega t}$ that we write as the product of a phase, which oscillates at the central frequency $\omega$ of the light pulse, and a complex amplitude $\vect E_\text{spp}(\vect r,t)$ that varies in time according to the envelope of the laser pulse and the propagation of the SPP wave. The SPP electric field measured by this technique at a delay time of $\Delta t=71.7$ fs is shown as a three dimensional rendering in Fig. \ref{fig:2}b. Applying Maxwell's equations to this measured field, we compute the SPP magnetic field $\vect H_\text{spp}(\vect r,t)e^{-i\omega t}$ (Fig. \ref{fig:2}c). This field lies entirely in the $x-y$ plane, which is the surface plane of the gold film. These fields are obtained at a maximal spatial resolution of $\approx$ 10 nm, which is the resolution of the PEEM, and with a delay-time-discretisation of 0.16 femtoseconds. We benchmark our new technique in Supplementary Note 5 by applying it to the trimeron spin texture that was previously investigated by Dai et al. \cite{Dai2020}
  
\subsection{Spin textures}
\begin{figure}[htp!]
\centering
\includegraphics[width=15.0cm]{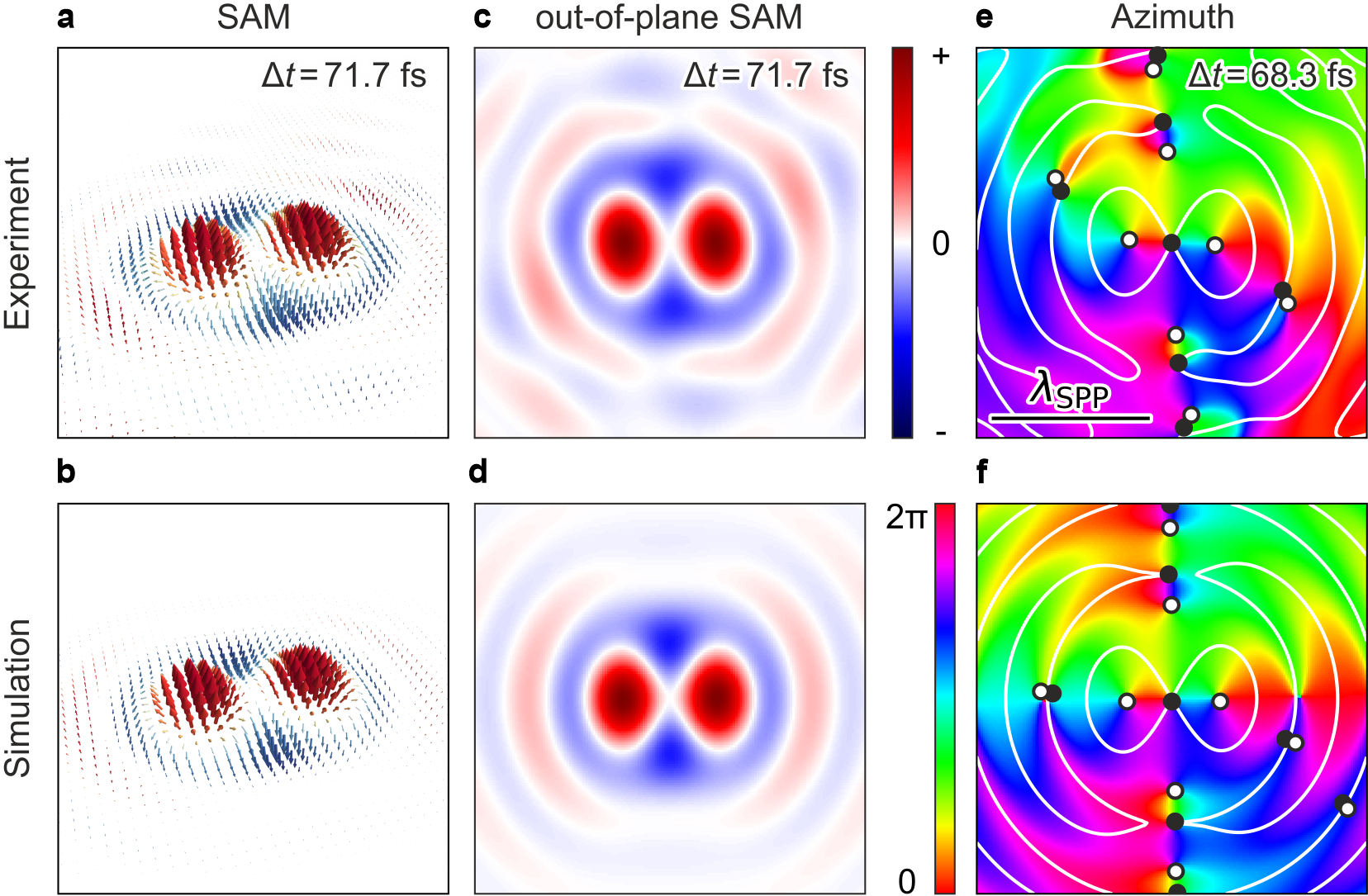}
\caption{A comparison between experiment and simulation for the spin vectors of the SPP meron pair. a,b) the spin vectors demonstrating the presence of the meron pair. The time-evolution of the experimentally determined spin vectors is shown in Video 3; c, d) the out-of-plane spin components $s_z$ where the colour encodes the spin direction. The spin is zero in the white regions that locate L-lines; e,f) the directions of in-plane components of the spin vectors are represented by the colour hue. The time-evolution of the experimentally determined azimuth is shown in Video 4. The white lines are the L-lines where the out-of-plane spin $s_z=0$.  The points where the in-plane vectors are zero (the zeroes of the field) are marked with dots, where white corresponds to zeroes with vorticity $v=+1$ and black for zeroes with vorticity $v=-1$. (Video 3, M4V, 5.8 MB; Video 4, M4V, 2.1 MB)}
\label{fig:3}
\end{figure}

From the complete measurements of the spatial and temporal evolution of the SPP electromagnetic fields we calculate the spin angular momentum density using
\begin{equation}
\label{eq:spin}
\vect s(\vect r,t)=\text{Im } \left (\epsilon \vect E_\text{spp}^*\times \vect E_\text{spp}+ \mu\vect H_\text{spp}^* \times \vect H_\text{spp} \right )/2\omega,
\end{equation}
as can be derived from the spin formula in Ref. \cite{Barnett2010} with appropriate substitutions for time harmonic fields. We show a three dimensional view of the vectors of this experimental spin field in Fig. \ref{fig:3}a which reveals two vortex quasi-particles, with spin vectors directed out of the plane, embedded in a region of weak downward directed spins. 

As we show below, each of these features has the local topology of a meron, and thus this quasi-particle corresponds to a \emph{meron pair}. From the measured time sequence of the SPP spin field, we find that the quasi-particle is stable for at least $23$ femtoseconds, as shown in Video 3, a value close to the overlap time of the pulsed SPP waves arriving from opposite sides of the Archimedean spiral. This stability arises despite the complex time and space variations of the underlying electromagnetic fields. The experimentally determined meron pair shows excellent agreement with the theoretical model (Fig. \ref{fig:3}b) based on a simulation of the SPP electric fields.\cite{Davis2017} Remarkably, the measured spin distribution of this SPP meron pair also shows qualitative agreement with simulated magnetic meron pairs expected in thin films\cite{Lu2020} but which have not yet been measured to such fine detail in magnetic systems. The out-of-plane spin components $s_z$ are shown for both experiment and simulation in Fig. \ref{fig:3}c-d. The white regions in these colour plots highlight the lines along which the out-of-plane spin component is zero $s_z=0$. These are the L-lines where the in-plane electric fields are linearly polarised and show no in-plane rotation.

An important topological feature of our measured vector field arises where the in-plane vectors are zero ($s_x=s_y=0$). Such `zeroes' of the in-plane field come in two types, one being a C-point\cite{Nye1983} where the spin field is oriented completely out of the plane ($s_z\neq 0$), and the other being an amplitude vortex\cite{Fosel2017} where the complete spin falls to zero ($s_x=s_y=s_z=0$). Both C-points and amplitude vortices of the spin field can be assigned a vorticity $v=\pm 1$ according to the direction of rotation of the vectors on a path about them, as depicted in Fig. \ref{fig:1}a . This property is shown by the change in the hue on traversing a path about a zero point, such as from blue to green or blue to red in Fig. \ref{fig:3}e for the experimentally determined spin fields and in Fig. \ref{fig:3}f for the numerical simulation. The zeroes of the field are highlighted by white dots ($v=+1$) and black dots ($v=-1$) in Figs. \ref{fig:3}e-f. It is clear that each of the merons in Fig. \ref{fig:3}a is associated with a C-point with a vorticity $v=1$ and a polarity $p=1/2$, since $s_z>0$ and therefore each has a Chern number of $C=p\cdot v=1/2$ or a total of $C=1$, as expected. Between the two merons the spin magnitude goes to zero (cf. Fig. \ref{fig:3}a,b) which corresponds to an amplitude vortex mid way between the C-points. This amplitude vortex has a vorticity of $v=-1$ and its presence is required by topology, as will be discussed below.

The vorticities associated with the zeroes of the in-plane spin field depicted in Fig. \ref{fig:3}e-f have a fundamental connection to topology such that vector fields with different vorticity are non-homotopic, i.e., they have a different topology.\cite{Simon2018}  The vorticity, when calculated around any zero, is also known as the Poincar\'e index. This index is positive or negative according to the direction of rotation of the field about the zero, exactly as shown for the vorticities in Fig. \ref{fig:1}a. Poincar\'e proved that the sum of the vorticities of a vector field is equal to the Euler characteristic $\chi$, which is a fundamental topological invariant. 

In our experiment, the SPPs created at each point on the Archimedean spiral have in-plane spin vectors parallel to the tangents of the grooves and therefore, on propagating over the surface, have the topology of a disk for which the Euler characteristic is $\chi=1$. As shown in Fig. \ref{fig:3}e-f, most of the zeroes come in pairs of opposite index, which therefore cancel. Since the meron pair involves two zeroes, each with vorticity $v=+1$, it is a topological necessity that there must be another zero with vorticity $v=-1$ such that the sum of vorticities is $\chi=1$. This is the origin of the amplitude vortex that lies between the two C-points of the meron pair. Based on the Euler charateristic we expect all other zeroes to appear in pairs of opposite index to maintain $\chi=1$. If an additional isolated vortex exists within the imaged field of view, a corresponding vortex with opposite vorticity must exist outside the field of view to fulfill the Poincar\'e-Hopf theorem. For the infinitely extending case of the meron spin fields there is an infinite number of zeroes from C-points and amplitude vortices whose indexes (except one) cancel pairwise. The measured out-of-plane components are essential for determining the polarity to distinguish between true spin vortices (C-points) and amplitude vortices. This highlights the importance of using polarimetric vector microscopy to measure the complete spin vector direction when using $C=p\cdot v$ to determine the Chern number: the in-plane components alone are insufficient to characterize the spin topology.  

The relationship between vorticity and the Euler characteristic implies a vortex conservation law in the two-dimensional vector field: vortices can only be created or annihilated in pairs with vorticities of opposite sign\cite{Simon2018,Dennis2009} such that the Euler characteristic remains constant. Since this is a constraint on the topology, the Poincar\'e sum becomes a global constraint on the in-plane components of the vector spin fields that impacts the local properties of their zeroes. The topological connection between local and global properties arises in other fields, as with defects in crystals,\cite{Mermin1979} for example screw dislocations where the local line defect represents a topological charge that impacts the crystal structure at large distances. In our time-dependent measures of spin texture, we observe such vortex pair creation and annihilation experimentally, which we show as Video 4.

\begin{figure}[htp!]
\centering
\includegraphics[width=15.0cm]{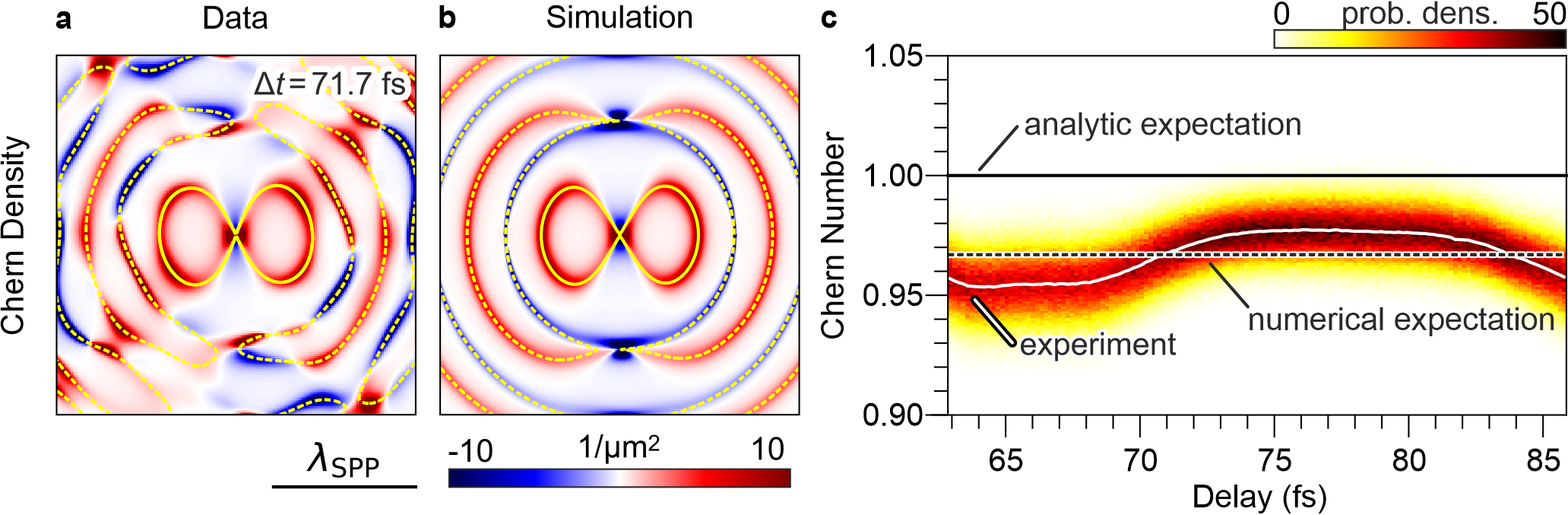}
\caption{The topology of the SPP meron pair. a) The experimental Chern density (kernel of Eq. \eqref{eq:1}) The time evolution of the experimental Chern density is shown in Video 5; b) the expected Chern density based on a numerical simulation of the SPP electric fields and their subsequent spin fields. The L-lines are marked in yellow; c) the Chern number within the marked solid L-line in (a) and (b) as a function of pump-probe delay. The experimental Chern number is shown as a white line, the result from a numerical simulation is shown as a dashed line and the theoretical value of $C=1$  is shown as a solid black line. The colored background represents the data reliability, based on a Monte-Carlo-based error propagation assuming a $10$ nm spatial resolution. (Video 5, M4V, 2.33 MB)}
\label{fig:4}
\end{figure}

From the measured spatio-temporal properties of the SPP spin field we calculate the Chern density, which is the kernel of the integral in Eq. \ref{eq:1}. The Chern density obtained from experimental spin data at a delay time of $\Delta t=71.7$ fs is shown in Fig. \ref{fig:4}a, with that calculated from a simulation shown for comparison in Fig. \ref{fig:4}b. The temporal dependence of the Chern density is shown in Video 5. The Chern number is the integral of the Chern density over a suitable region of the SPP field. However, to correctly measure the local topology, we need to limit the integral to the area close to the meron pair. The boundary of the integral is chosen as the L-line surrounding the spin feature,\cite{Dai2020} which is defined by the points where $s_z=0$, since the in-plane electric field vectors on this line are perfectly linearly polarised. The L-lines in the measured data are marked yellow in Fig. \ref{fig:4}a. With the experimentally determined SPP spin in Eq. \eqref{eq:1} we obtain the time-dependent Chern number shown in Fig.\ref{fig:4}c. The experimental data gives slightly varying numbers that strongly depend on the chosen L-line integration contour. Adding noise to the positioning of the contour in a Monte-Carlo-based error propagation, using the electron-optical resolution of the microscope of $10$ nm, results in experimental Chern numbers within the color-coded probability density in the background in Fig. \ref{fig:4}c. The temporal dependence of the Chern number shows that the meron pair remains topologically stable for the period of time where the SPP waves from the boundary overlap and interfere at the centre of the sample.

We note that the theoretical expectation is a Chern number of 1 for the meron pair, since each meron contributes a Chern number of $1/2$. The Chern number obtained from a temporal average of the experimental data is $C=0.97$. This number is smaller than $C=1$ due to the infinitely small central amplitude vortex and the finite pixelation of the data. How the Chern number depends on the resolution and  pixelation is discussed in detail in Supplementary Note 6. The dependence of the Chern number on the resolution emphasizes the importance of gathering high resolution spatial measurements of the SPP field in order to extract the underlying topology and proper Chern numbers. A pixelation of 12 nm, which corresponds to the pixelation by the electron detector, is expected to yield a Chern number of $C=0.97$ (cf. the numerical expectation in Fig.~\ref{fig:4}c).

\section*{Conclusion and outlook}
In our analysis we have identified regions in the SPP spin field associated with C-points and L-lines and unambiguously identified a meron pair. It is worth remarking that these are equivalent to the topology of a landscape as Maxwell also investigated,\cite{Maxwell1870} where the contour lines and their relationship to points representing maxima (hills) or minima (dales) appear, respectively, like the L-lines and C-points here. The determination of the SPP spin topology was only possible using our polarimetric PEEM method that enables the measurement of the three-dimensional SPP vector fields and their evolution in time. The topology of these SPP vector fields is generally hidden in the static field distributions obtained from interferometric optical near field measurements, and the presence of amplitude vortices can further complicate the analysis.

The time and space resolution that we obtain provides opportunities for studying other complex vector wave interference phenomena dictated by an underlying topology, such as wave patterns associated with two-dimensional quasicrystals.\cite{Vardeny2013} Although we have demonstrated polarimetric PEEM with long-range surface plasmons, the method is equally as well suited to studying short range surface plasmons. These short range SPPs have potential as highly compact localized field sources with wavelengths an order of magnitude smaller than the excitations in free space\cite{Du2019,Frank2017,Shen2021}. Such fields can be intense yielding possibilities for highly localized excitation and interactions with potential applications in structured light illumination and microscopy, as well as in complex light field-matter interaction. Topology then becomes an important tool for controlling near fields at the nanoscale because topological protection can help to stabilize the field geometries.\cite{Xue2021}

\section*{Author Contributions}

D.J., P.D., F.MzH, A.N., A.R. performed the time-resolved PEEM experiments. B.F. and T.M. prepared the samples, D.J., P.D., F.MzH, H.G. conceived the experiment, D.J., P.D., A.N., T.D., F.MzH analyzed the data, T.D., D.J., P.D., F.MzH, H.G. wrote the manuscript with contributions from all authors. D.J, P.D and A.N. contributed equally and should be considered co‑first authors.

\section*{Data availability statement}
The data used for the analysis will be made available upon reasonable request.

\section*{Competing Interest statement}
The authors declare no competing interest.

\section*{Acknowledgements}
The authors acknowledge support from the ERC (Complexplas, 3DPrintedoptics), DFG (SPP1391 Ultrafast Nanooptics, CRC 1242 “Non-Equilibrium Dynamics of Condensed Matter in the Time Domain” project no. 278162697-SFB 1242), BMBF (Printoptics), BW Stiftung (Spitzenforschung, Opterial), Carl-Zeiss Stiftung. T.J.D. acknowledges support from the MPI Guest Professorship Program and from the DFG (GRK2642) Photonic Quantum Engineers for a Mercator Fellowship. We acknowledge fruitful discussions about topology with Karin Everschor-Sitte and David Paganin for alerting us to Maxwell's paper on topology.

\bibliographystyle{unsrt}
\bibliography{meron}

\end{document}